\documentclass[journal]{IEEEtran}

\ifCLASSINFOpdf
  
\else

\fi

\usepackage{algorithm}
\usepackage{algorithmic}
\usepackage{amssymb}
\usepackage{nccmath}
\usepackage{cite}
\usepackage{graphicx}  
\usepackage[table]{xcolor}
\usepackage{verbatim}

\usepackage[utf8]{inputenc}
\usepackage{amsmath}
\usepackage{amsthm}     
\usepackage{accents}

\newcommand{\ubar}[1]{\underaccent{\bar}{#1}}
\begin{document}

\title{Robust Optimal Energy and Reserve Management of Multiple-Microgrids via Cooperative Bidding}%
\author{L.~P.~M.~I.~Sampath,~Ashok~Krishnan, Y.~S.~Foo~Eddy,~\IEEEmembership{Member,~IEEE}
and~H.~B.~Gooi,~\IEEEmembership{Senior~Member,~IEEE \vspace{-6mm}}
\thanks{L. P. M. I. Sampath is with the Interdisciplinary Graduate School, Nanyang Technological University, Singapore 637371 (e-mail: mohashai001@e.ntu.edu.sg).} 
\thanks{Ashok Krishnan, Y. S. Foo Eddy and H. B. Gooi are with the School of Electrical and Electronic Engineering, Nanyang Technological University, Singapore 639798. 
}}%


\maketitle


\begin{abstract}
This work proposes a cooperative trading scheme for the robust optimal energy and reserve management in a multiple-microgrid (MMG) system comprising four microgrids (MGs). This scheme includes a robust optimization (RO) model which accounts for the uncertainties in the renewable energy source (RES) generation, the load demand and the energy market prices using a scenario based worst-case exploration technique. A coordinating entity in the MMG system called a power sharing operator manages the day-ahead energy and reserve trading between the MGs and also with the main grid, thereby enabling market clearing within the MMG system. The proposed scheme is based on the iterative sub-gradient method.
The results showcase the convergence characteristics of the proposed scheme. Moreover, the tradeoff between the energy and reserve costs in each MG are also analysed. Finally, the economic benefits gained by the MMG system under the proposed RO based cooperative trading scheme are evaluated.

\end{abstract}

\begin{IEEEkeywords}
Cooperative trading, distributed optimization, iterative sub-gradient method, multiple-microgrids, optimal dispatch, robust optimization.
\end{IEEEkeywords}

\IEEEpeerreviewmaketitle
\vspace{-5mm}
\section{Introduction}
\IEEEPARstart{M}{icrogrids} are gaining popularity due to their ability to integrate RESs into the electrical power system. 
In addition to RESs, a typical MG may include controllable sources such as diesel generators (DGs) and energy storage systems (ESSs) apart from critical and dispatchable (price-responsive) loads. Furthermore, the increasing levels of RES penetration, advancements in communication technologies, and the liberalization of energy markets have supported the proliferation of MMG systems. The sharing of resources between the MGs constituting an MMG system has been an area of active research in recent years~\cite{Nguyen18,Mohasha}. In this context, an earlier work proposed a multi-agent based framework to describe the concept of cooperative MMG operation~\cite{EJNg10}. More recently, the merits of sharing resources between the MGs constituting an MMG system using a cooperative framework were espoused in \cite{Mohasha,YunLiu18,YZLi18,KRahbar18}.
In every MG, a microgrid operator (MGO) utilizes the RESs and optimally dispatches the controllable sources and flexible loads to satisfy the nominal demand. Two major sources of uncertainty affect the economic and reliable operation of MGs. Firstly, the RES generation and the load demand are stochastic and non-dispatchable. Consequently, an arbitrary allocation of reserves to support the demand and RES uncertainties may result in suboptimal or unreliable MG operation~\cite{YSasaki18}. Secondly, in an energy market (for instance, the National Electricity Market of Singapore), the prices at which electricity is purchased from and sold to the main grid are allowed to change up to $ 65 $ minutes prior to the actual trade \cite{YunLiu18,YZLi18,Dante18}. 
%

%

To mitigate these uncertainties efficiently, the energy and reserve within a MG must be jointly dispatched in an optimal fashion. To enable this, numerous stochastic optimization based models have been proposed in the literature. 
For example, a stochastic MG scheduling model was developed in~\cite{Su14} to minimize the expected operating cost of a MG with intermittent RESs. Furthermore, the bidding strategy proposed by the authors of~\cite{GLiu16} used a hybrid stochastic/robust optimization model in a centralized architecture for operating grid-connected MGs. However, stochastic modeling requires the probability distributions of random factors to be known in advance. As shown in~\cite{GLiu16}, this information is hard to obtain for the energy market prices. Conversely, RO procedures characterize uncertainties using confidence bounds which can be computed using historical data and statistical inference techniques~\cite{Pinson10}. In general, RO procedures minimize the total cost while providing a feasible solution under the worst case scenario which is generated on the basis of the uncertainties present in the optimization model \cite{YuZhang13,YunLiu18,YZLi18,Xiang16}. Reference~\cite{Ding16} focused on the robust co-optimization of energy and reserves in real-time energy markets.
A RO model with a polyhedral uncertainty set was used in~\cite{YuZhang13} to incorporate a vertex enumerating algorithm for worst-case exploration. However, exploring all the vertices of the uncertainty polyhedron as proposed in \cite{YuZhang13} is computationally expensive due to an exponential rise in the number of vertices with a rise in the number of uncertain variables.
Conversely, \cite{Xiang16} 
adopted a scenario-based RO framework in which the \emph{Taguchi's orthogonal array testing} (TOAT) method was used to select a fraction of the vertices from the uncertainty polyhedron. The TOAT method is a relatively scalable and practical approach towards searching for the worst-case scenario in the operation of the MG. 
 %

In the context of MMG systems, a coordinating entity may be used for interconnecting the constituent MGs to enable power sharing~\cite{YZLi18}. This framework allows individual MGOs to optimize their local resources and cooperatively contribute to the MMG energy and reserve markets via the coordinating entity. Furthermore, this framework is also privacy-preserving which only requires the individual MGOs to communicate bidding (non-critical) information \cite{YunLiu18,YZLi18,KRahbar18,Gregoratti15}. The optimal joint day-ahead dispatch of energy and allocation of reserves using RO based approaches for MMG systems were recently proposed in \cite{YunLiu18} and \cite{YZLi18}. To facilitate the clearing of the MMG energy and reserve markets, an alternating direction method of multipliers based approach was used in \cite{YunLiu18} while a game theoretic approach was investigated in \cite{YZLi18}. Reference \cite{YunLiu18} only considered the sharing of energy between MGs while managing reserves locally within the MG, thereby diluting the benefits offered by the MMG framework while also resulting in a possibly suboptimal dispatch solution. Reference \cite{YZLi18} also facilitated the sharing of reserves among the constituent MGs in an MMG system. However, in \cite{YZLi18}, the realized worst case scenario assumes the lowest RES output and the highest demand (and vice-versa) simultaneously over the entire optimization horizon for all the constituent MGs. This scenario is highly unlikely in practice. Furthermore, the importance of considering the ramping capabilities of the generators (such as DGs and ESSs) while allocating reserves was highlighted in~\cite{German16,YSasaki18,Nguyen18}. However, this has been ignored by the existing works including \cite{YZLi18,YunLiu18}.
\vspace{-2mm}
\subsection{Contributions}
This paper proposes a detailed model for the day-ahead joint optimal dispatch of the energy and allocation of the reserves in a MMG system comprising four MGs which are interconnected through a coordinating entity called the power sharing operator (PSO). For each MG, the TOAT method is used to generate a reduced number of uncertain scenarios, each of which represents a combination of vertices of the polyhedral uncertainty set. This polyhedral set is characterized by the uncertainties in the RES generation, the load demand and the energy market prices. Based on the earlier discussions and compared with the existing works in the literature, the salient contributions of this paper are enumerated below:
\begin{enumerate}
	\item A cooperative bidding scheme based on the iterative sub-gradient method in \cite{Gregoratti15} is proposed to facilitate the trading of energy and reserves between the constituent MGs. The proof of convergence for this scheme to the centralized solution is also provided.
	\item The reserve allocation is performed by respecting the intertemporal ramping capabilities of the DGs and the ESSs in the constituent MGs. This facilitates the robust dynamic feasibility of the energy dispatch and the reserve allocations.
	\item The RO model also includes the participation of flexible loads in the energy dispatch and reserve allocation.
\end{enumerate}
The results demonstrate the benefits of sharing energy and reserves between MGs when compared with the isolated operation of individual MGs. The performance of the proposed RO model is also compared with a deterministic optimization model.
%
\vspace{-3mm}%
\section{Uncertainty Model for Robust Optimization}
\subsection{Uncertainty Characterization}
The \emph{actual} values of the demand and the RES power outputs are not known \textit{a priori}. Several forecasting techniques use historical datasets to predict the demand pattern and the RES generation. However, the accuracy of such forecasts is not guaranteed. Some forecasting techniques utilize probability distributions while others use time series methods. In this paper, it is assumed that the uncertain variables lie within a polyhedral set. This is a general approach for modelling uncertainties which only requires the deterministic values for the upper and lower bounds of each uncertain variable over time. These confidence bounds can be computed based on historical data using statistical inference techniques~\cite{Pinson10}. This approach also determines the worst-case scenario for the RO problem solved in Section~\ref{Rob_Opt}.

\vspace{-2mm}
\subsection{Worst-Case Exploration Techniques}
To obtain feasible results under the worst-case scenario, the cost minimum resource management problem needs to be \emph{maximized} over the polyhedral uncertainty set. The resulting \emph{max-min} problem is a nonconvex problem which is NP-hard to solve in polynomial time. However, the specific structure of this problem enables the global solution to be found at the extreme points of the uncertainty polytope~\cite{YuZhang13}. The global solution can be obtained if the resource optimization problem is evaluated at all the vertices of the uncertainty polytope. However, the computational complexity involved in exploring all the vertex (full factorial) combinations increases exponentially with the dimensions of the uncertainty set \cite{YuZhang13}. Thus, there is a need for alternative tractable solution approaches. %

In this context, the TOAT method generates a special set of \emph{orthogonal arrays} (OAs) which represents a fraction of the full factorial combinations of experiments. The general representation of an OA is $ L_HB^F $, where $ H $ is the number of experiments/scenarios; $ F $ is the number of random factors and $ B $ is the number of levels (the two confidence interval bounds) considered for each random factor. The TOAT method combines the least possible scenarios which are uniformly distributed over the uncertainty space. Thus, the selected uncertainty set has the potential to represent the worst-case scenario~\cite{Wang17,Xiang16}. Interested readers may refer to~\cite{TOAT} for a detailed review of the applications of the TOAT method in power flow studies.\vspace{-2mm}%

\begin{table}[ht!]
	\centering
	\vspace{-3mm}
	\caption{\footnotesize Selected Scenarios based on the OA $ L_82^7 $}
	\label{OA}
	\vspace{-3mm}
	\begin{tabular}{c|c c c c c}
		\hline
		Scenario & \multicolumn{5}{c}{Levels of each factor}  \\  \cline{2-6}
		$ (h) $	& PSO & MG $ 1 $ & MG $ 2 $ & MG $ 3 $ & MG $ 4 $\\
		\hline \hline
		$ 1 $ & $ +1 $ & $ +1 $ & $ +1 $ & $ +1 $ & $ +1 $\\
		$ 2 $ & $ +1 $ & $ +1 $ & $ +1 $ & $ -1 $ & $ -1 $\\
		$ 3 $ & $ +1 $ & $ -1 $ & $ -1 $ & $ +1 $ & $ +1 $\\
		$ 4 $ & $ +1 $ & $ -1 $ & $ -1 $ & $ -1 $ & $ -1 $\\
		$ 5 $ & $ -1 $ & $ +1 $ & $ -1 $ & $ +1 $ & $ -1 $\\
		$ 6 $ & $ -1 $ & $ +1 $ & $ -1 $ & $ -1 $ & $ +1 $\\
		$ 7 $ & $ -1 $ & $ -1 $ & $ +1 $ & $ +1 $ & $ -1 $\\
		$ 8 $ & $ -1 $ & $ -1 $ & $ +1 $ & $ -1 $ & $ +1 $\\
		\hline
	\end{tabular}\vspace{-2mm}
\end{table}
Let there be an MMG with four MGs. Table~\ref{OA} depicts the first five columns of the OA $ L_82^7 $ specifying eight experiments/scenarios involving five two-level factors (five cooperative entities)~\cite{TOAT}. The reserve requirement of a MG can be either positive or negative depending on the variations in the demand and RES generation. The $ +1 $ and $ -1 $ levels represent the positive and negative reserve requirements respectively. In this context, the expected levels of the nominal demand can be defined as shown in~\eqref{P_ND}.
Let $ \mathcal{H}=\{0,1,\ldots,H\} $ be the set of scenarios, wherein the base-case is represented by $ h=0 $ and let $ \mathcal{T}=\{1,\ldots,T\} $ be the optimization horizon of the study. The superscripts $ m $ and $ t $ denote the MG number and the optimization interval respectively associated with the corresponding variable or parameter. 
\begin{subequations}\label{P_ND}%
	\begin{align}%
	\left[ {P}^{m,t}_{{\rm ND}}\right]_+:= {P}^{m,t}_{{\rm ND},0}  +P^{m,t,\delta+}_{\rm ND} \\
	\left[ {P}^{m,t}_{{\rm ND}}\right]_-:= {P}^{m,t}_{{\rm ND},0} -P^{m,t,\delta-}_{\rm ND}
	\vspace{-5mm}
	\end{align}%
\end{subequations}%
where nonnegative parameters $ {P}^{m,t}_{{\rm ND},0},\ P^{m,t,\delta+}_{\rm ND}$ and $ P^{m,t,\delta-}_{\rm ND} $ are the mean, and the deviations of the upper and lower confidence bounds of the nominal demand of MG $ m $ during hour $ t $, respectively. Then, the confidence interval of the nominal demand $ P^{m,t}_{{\rm ND},h} $ of MG $ m $ during hour $ t $ can be defined as follows:\vspace{-1mm}%
\begin{equation}\label{key}
P^{m,t}_{{\rm ND},h} \in \Big\{\left[ {P}^{m,t}_{{\rm ND}}\right]_+,\ \left[ {P}^{m,t}_{{\rm ND}}\right]_-\Big\};\ \forall h \in \mathcal{H} \backslash \{0\}
\end{equation}
Furthermore, the expected levels of the wind power generation can be defined as follows:
\begin{subequations}\label{P_WE}
	\begin{align}
	\left[ P^{m,t}_{{\rm W},w}\right]_+ :={P}^{m,t}_{{\rm W},w,0}-P^{m,t,\delta-}_{{\rm W},w} \\
	\left[ P^{m,t}_{{\rm W},w}\right]_- :={P}^{m,t}_{{\rm W},w,0}+P^{m,t,\delta+}_{{\rm W},w}
	\end{align}
\end{subequations}
The expected levels of the solar photovoltaic (PV) power generation $ \big[ P^{m,t}_{{\rm PV},s}\big]_+ $ and $ \big[ P^{m,t}_{{\rm PV},s}\big]_- $ can also be defined in a similar manner. For the PSO, the $ +1 $ and $ -1 $ levels represent the upper and lower bounds of the price for purchasing electricity from the main grid respectively.
\begin{subequations}\label{grid_price}
	\begin{align}
	\left[ \gamma^{t}_{{\rm b}}\right]_+ :={\gamma}^{t}_{{\rm b},0} +\gamma^{t,\delta+}_{{\rm b}} \\
	\left[ \gamma^{t}_{{\rm b}}\right]_- :={\gamma}^{t}_{{\rm b},0} -\gamma^{t,\delta-}_{{\rm b}}
	\end{align}
\end{subequations} 
The expected levels of the price for selling electricity to the main grid (represented by $ \left[ \gamma^{t}_{{\rm s}}\right]_+ $ and $ \left[ \gamma^{t}_{{\rm s}}\right]_- $) can also be defined in a similar fashion.
\vspace{-3mm}%
\section{Robust Day-Ahead Optimal Energy Management of Multi-Microgrids}\label{Rob_Opt}
\subsection{Multi-Microgrids System Structure}
An MMG system is a cluster of MGs which are individually connected to a common bus called the point-of-common coupling (PCC) via a circuit breaker, for instance~\cite[Fig.~1]{YZLi18}. Each MG in the MMG system can operate either in the islanded or the grid-connected mode.
The PCC is connected to the main grid. The PSO manages the import and export of electricity through the PCC at predetermined buying and selling prices respectively. In this context, it is assumed that the supply and demand balance of the MMG system is maintained at all times under the grid-connected mode. Each MG in the MMG system comprises components such as DGs, dispatchable loads, fixed loads, ESSs and RESs. The MGOs need to optimally meet the local MG demand without violating any technical constraints.
Based on the local demand-supply scenario, each MG may seek to import/export electricity from/to the PSO at different times in a day. The PSO clears the MMG electricity market based on the supply and demand offers received from the constituent MGs. The PSO enables the optimal operation of the entire MMG system by facilitating power sharing among the interconnected constituent MGs via a cooperative bidding scheme, the details of which are presented in the later sections.

\vspace{-3mm}
\subsection{Microgrid Component Modeling}
The MG energy management system usually solves an optimal day-ahead dispatch problem which is subject to the power balance constraint of the MG and the operational constraints of the MG components. To enable the formulation of this dispatch problem, the cost functions and the operational constraints of all the MG components are developed in the following paragraphs. The MG components are modelled to enable the robust solution of the optimal dispatch problem under all the simulated uncertain scenarios (including the worst-case).
The superscripts `$ {\rm max} $' and `$ {\rm min} $' represent the upper and lower bounds respectively of the corresponding variable or parameter. 
\subsubsection{Diesel Generators (DGs)}
Let $ \mathcal{G}^m $ be the set of DGs in MG~$ m $. For all $ g \in \mathcal{G}^m  $:
\begin{subequations}
	\begin{align}
	&C^{m,t}_{{\rm G},g,h}\left(P^{m,t}_{{\rm G},g,h}\right)=a^m_{{\rm G},g}\left(P^{m,t}_{{\rm G},g,h}\right)^2+b^m_{{\rm G},g}P^{m,t}_{{\rm G},g,h}\label{Cost_DG}\\
	&(P^{m}_{{\rm G},g})^{\rm min} \leq P^{m,t}_{{\rm G},g,h} \leq (P^{m}_{{\rm G},g})^{\rm max};\, \forall h \in \mathcal{H}
	\label{Cap_DG1}\\
	& -r^{{\rm dn},m}_{{\rm G},g} \leq P^{m,t}_{{\rm G},g,h}-P^{m,t}_{{\rm G},g,0} \leq r^{{\rm up},m}_{{\rm G},g};\,\forall h \in \mathcal{H}\backslash\{0\}
	\label{ResG}\\
	&\max_{h \in \mathcal{H}}P^{m,t}_{{\rm G},g,h}-\min_{h \in \mathcal{H}}P^{m,t-1 }_{{\rm G},g,h} \leq R^{{\rm up},m}_{{\rm G},g}
	\label{Ramp_DG2}\\
	&\max_{h \in \mathcal{H}}P^{m,t-1 }_{{\rm G},g,h} - \min_{h \in \mathcal{H}}P^{m,t}_{{\rm G},g,h} \leq R^{{\rm dn},m}_{{\rm G},g}
	\label{Ramp_DG1}
	\end{align}
\end{subequations}
where $ P^{m,t}_{{\rm G},g,h} $ is the power output of DG~$ g $ under scenario~$ h $; $ a^m_{{\rm G},g} $ and $ b^m_{{\rm G},g} $ are the quadratic and linear coefficients respectively of the fuel cost function~\eqref{Cost_DG}; $ r^{{\rm up},m}_{{\rm G},g} $ and $ r^{{\rm dn},m}_{{\rm G},g} $ are the up and down reserve capability limits respectively of DG $g$; $ R^{{\rm up},m}_{{\rm G},g} $ and $ R^{{\rm dn},m}_{{\rm G},g} $ are the up and down ramp rate limits respectively of DG $g$. Equation~\eqref{Cap_DG1} represents the generation bounds of DG $g$ under scenario~$ h $. 
Constraint~\eqref{ResG} ensures that the power dispatch under scenario $ h $ respects the reserve capability bounds. 
Constraints \eqref{Ramp_DG2} and \eqref{Ramp_DG1} ensure the robust feasibility of the intertemporal up and down ramp rate constraints respectively under scenario $ h $. %
Constraints \eqref{Ramp_DG2} and \eqref{Ramp_DG1} are nonconvex in nature. These constraints are linearized as follows using the additional decision variables $ \bar{P}^{m,t}_{{\rm G},g} $ and $ \ubar{P}^{m,t}_{{\rm G},g} $. For all $ g \in \mathcal{G}^m $:
\begin{subequations}\label{lin_DG}
	\begin{align}
	&\bar{P}^{m,t}_{{\rm G},g}\geq {P}^{m,t}_{{\rm G},g,h},\ \ubar{P}^{m,t}_{{\rm G},g}\leq {P}^{m,t}_{{\rm G},g,h};\ \forall h \in \mathcal{H}
	\\
	&\bar{P}^{m,t}_{{\rm G},g}-\ubar{P}^{m,t-1}_{{\rm G},g} \leq  R^{{\rm up},m}_{{\rm G},g}
	\\
	&\bar{P}^{m,t-1}_{{\rm G},g} -\ubar{P}^{m,t}_{{\rm G},g} \leq R^{{\rm dn},m}_{{\rm G},g}
	\end{align}
\end{subequations}
\subsubsection{Energy Storage System (ESS)}%
Let $ \mathcal{E}^m $ be the set of ESSs in MG~$ m $. For all $ e \in \mathcal{E}^m  $:
\begin{subequations}
	\begin{align}
	&C^{m,t}_{{\rm E},e,h}\left(P^{{\rm c},m,t}_{{\rm E},e,h},P^{{\rm d},m,t}_{{\rm E},e,h}\right)=a^m_{{\rm E},e}\left(P^{{\rm c},m,t}_{{\rm E},e,h}+P^{{\rm d},m,t}_{{\rm E},e,h}\right)^2 \label{Cost_ESS}\\
	&0 \leq P^{{\rm c},m,t}_{{\rm E},e,h} \leq (P^{{\rm c},m}_{{\rm E},e})^{\rm max};\, \forall h \in \mathcal{H}
	\label{Ech1}\\
	&0 \leq P^{{\rm d},m,t}_{{\rm E},e,h} \leq (P^{{\rm d},m}_{{\rm E},e})^{\rm max};\, \forall h \in \mathcal{H}
	\label{Edch1}\\
	&E^{m,t}_{e,0}=E^{m,t-\Delta t}_{e,0}+\frac{\Delta t}{B^m_e}\left(\eta^{\rm c}_{e}P^{{\rm c},m,t}_{{\rm E},e,0}-\frac{1}{\eta^{\rm d}_{e}}P^{{\rm d},m,t}_{{\rm E},e,0}\right)
	\label{SOC}\\
	&\max_{h \in \mathcal{H}}E^{m,t}_{e,h} \geq \max_{h \in \mathcal{H}}E^{m,t-\Delta t}_{e,h}\nonumber \\
	&\hspace{1cm} +\frac{\Delta t}{B^m_e}\left(\eta^{\rm c}_{e}\max_{h \in \mathcal{H}}P^{{\rm c},m,t}_{{\rm E},e,h}-\frac{1}{\eta^{\rm d}_{e}}\min_{h \in \mathcal{H}}P^{{\rm d},m,t}_{{\rm E},e,h}\right)
	\label{SOC2}\\
	&\min_{h \in \mathcal{H}}E^{m,t}_{e,h} \leq \min_{h \in \mathcal{H}}E^{m,t-\Delta t}_{e,h}\nonumber \\
	&\hspace{1cm}+\frac{\Delta t}{B^m_e}\left(\eta^{\rm c}_{e}\min_{h \in \mathcal{H}}P^{{\rm c},m,t}_{{\rm E},e,h}-\frac{1}{\eta^{\rm d}_{e}}\max_{h \in \mathcal{H}}P^{{\rm d},m,t}_{{\rm E},e,h}\right)
	\label{SOC1}\\
	&(E^{m}_{e})^{\rm min} \leq E^{m,t}_{e,h} \leq (E^{m}_{e})^{\rm max};\, \forall h \in \mathcal{H}
	\label{Cap_E2}\\
	&E^{m,T}_{e,0}={E}^{m}_{{\rm R},e}
	\label{E0}\\
	& 0.8{E}^{m}_{{\rm R},e} \leq {E}^{m,T}_{e,h} \leq 1.2{E}^{m}_{{\rm R},e};\, \forall h \in \mathcal{H} \backslash \{0\}
	\label{E01}
	\end{align}
\end{subequations}
where $ P^{{\rm c},m,t}_{{\rm E},e,h} $, $ P^{{\rm d},m,t}_{{\rm E},e,h} $, $ E^{m,t}_{e,h} $ and $ B^m_e $ are the charging power, the discharging power, the state-of-charge (SOC) and the energy capacity respectively of ESS $ e $ under scenario~$ h $ and $ a^m_{{\rm E},e} $ is the coefficient of the quadratic degradation cost function which is included in~\eqref{Cost_ESS} of ESS $ e $. Finally, $ \eta^{\rm c}_{e} $ and $ \eta^{\rm d}_{e} $ are the charging and discharging efficiencies of ESS $ e $ respectively. The charging and discharging power limits of ESS $ e $ are provided in \eqref{Ech1} and \eqref{Edch1} respectively. The robust feasibility of the time varying SOC of ESS $e$ within its operational bounds is ensured by \eqref{SOC}$ - $\eqref{Cap_E2}; $ \Delta t $ is the length of an interval in the optimization period (\emph{i.e.}, $ 1\,{\rm h} $). Constraints~\eqref{E0}$ - $\eqref{E01} allow sufficient flexibility for the ESS operation during the following day, wherein $ {E}^{m}_{{\rm R},e} $ is the reference SOC (normally the initial SOC). However, \eqref{SOC2} and \eqref{SOC1} are nonconvex constraints which are linearized using additional decision variables for the SOC, the charging power and the discharging power as shown below in~\eqref{lin_ESS}. For all $ e \in \mathcal{E}^m $:
\begin{subequations}\label{lin_ESS}
	\begin{align}
	&\bar{E}^{m,t}_{e}\geq E^{m,t}_{e,h},\ \ubar{E}^{m,t}_{e}\leq E^{m,t}_{e,h}, \nonumber \\ 
	&\bar{P}^{{\rm c},m,t}_{{\rm E},e,h} \geq P^{{\rm c},m,t}_{{\rm E},e,h},\ \ubar{P}^{{\rm c},m,t}_{{\rm E},e,h} \leq P^{{\rm c},m,t}_{{\rm E},e,h}, \nonumber \\
	&\bar{P}^{{\rm d},m,t}_{{\rm E},e,h} \geq P^{{\rm d},m,t}_{{\rm E},e,h},\ \ubar{P}^{{\rm d},m,t}_{{\rm E},e,h} \leq P^{{\rm d},m,t}_{{\rm E},e,h};\ \forall h \in \mathcal{H}
	\\
	&\bar{E}^{m,t}_{e} \geq \bar{E}^{m,t-\Delta t}_{e}+\frac{\Delta t}{B^m_e}\left(\eta^{\rm c}_{e}\bar{P}^{{\rm c},m,t}_{{\rm E},e}-\frac{1}{\eta^{\rm d}_{e}}\ubar{P}^{{\rm d},m,t}_{{\rm E},e}\right)
	\\
	&\ubar{E}^{m,t}_{e} \leq \ubar{E}^{m,t-\Delta t}_{e}+\frac{\Delta t}{B^m_e}\left(\eta^{\rm c}_{e}\ubar{P}^{{\rm c},m,t}_{{\rm E},e}-\frac{1}{\eta^{\rm d}_{e}}\bar{P}^{{\rm d},m,t}_{{\rm E},e}\right)
	\end{align}
\end{subequations}

\subsubsection{Dispatchable Loads}
The MGs include critical loads which need to be strictly satisfied. Furthermore, the MGs support the demand response scheme formulated below.
\begin{subequations}
	\begin{align}
	&C^{m,t}_{{\rm FD},h}\left(P^{m,t}_{{\rm RD},h},P^{m,t}_{{\rm CD},h}\right)=a^m_{\rm FD}\left(P^{m,t}_{{\rm RD},h}+P^{m,t}_{{\rm CD},h}\right)^2 \label{Cost_D}\\
	&P^{m,t}_{{\rm D},h}=P^{m,t}_{{\rm RD},h}-P^{m,t}_{{\rm CD},h}+P^{m,t}_{{\rm ND},h};\ \forall h \in \mathcal{H} \label{D}\\
	&0 \leq P^{m,t}_{{\rm RD},h} \leq (P^{m,t}_{\rm RD})^{\rm max};\ \forall h \in \mathcal{H} \label{Cap_RD}\\
	&0 \leq P^{m,t}_{{\rm CD},h} \leq (P^{m,t}_{\rm CD})^{\rm max};\ \forall h \in \mathcal{H} \label{Cap_CD}\\
	&\sum_{t \in \mathcal{T}} \left[  {P}^{m,t}_{{\rm RD},0}-{P}^{m,t}_{{\rm CD},0} \right]  = 0 \label{FD}\\
	&\Delta t\sum_{t \in \mathcal{T}} \left[ \max_{h \in \mathcal{H}}{P}^{m,t}_{{\rm CD},h}-\min_{h \in \mathcal{H}}{P}^{m,t}_{{\rm RD},h}\right]  - E^m_{\rm Shed}\leq 0 \label{LS}
	\end{align}
\end{subequations}
where $ P^{m,t}_{{\rm ND},h} $, $ P^{m,t}_{{\rm CD},h} $, $ P^{m,t}_{{\rm RD},h} $ and $ P^{m,t}_{{\rm D},h} $ are the nominal (critical), curtailed, redispatched and actual demands under scenario~$ h $ respectively; and $ a^m_{\rm FD} $ is the coefficient of the cost function~\eqref{Cost_D} which reflects the sensitivity of the inconvenience felt by the consumers due to the load shifting. The actual demand in~\eqref{D} is considered to be the nominal demand after the flexibility adjustment. The demand redispatch and curtailment are constrained by \eqref{Cap_RD} and \eqref{Cap_CD} respectively. There is no load shedding permitted under the base-case scenario as described in~\eqref{FD}. Equation ~\eqref{LS} restricts the total load shedding under the worst-case scenario to $ E^m_{\rm Shed} $ and is linearized using additional decision variables $ \ubar{P}^{m,t}_{\rm RD} $ and $ \bar{P}^{m,t}_{\rm CD} $ for the redispatched and curtailed demands respectively as shown below.
\begin{subequations}\label{lin_D}
	\begin{align}
	&\ubar{P}^{m,t}_{\rm RD} \leq P^{m,t}_{{\rm RD},h},\ \bar{P}^{m,t}_{\rm CD} \geq P^{m,t}_{{\rm CD},h};\ \forall h \in \mathcal{H}\\
	&0 \leq \ubar{P}^{m,t}_{\rm RD} \leq (P^{m,t}_{\rm RD})^{\rm max},\ 0 \leq \bar{P}^{m,t}_{\rm CD} \leq (P^{m,t}_{\rm CD})^{\rm max}\\
	&\Delta t \sum_{t \in \mathcal{T}} \left[ \bar{P}^{m,t}_{\rm CD}-\ubar{P}^{m,t}_{\rm RD}\right]  \leq E^m_{\rm Shed}
	\end{align}
\end{subequations}
\subsubsection{Interactions with the PSO}
The PSO supervises the power and price bids submitted by each MGO for buying/selling electricity. The PSO first tries to optimally match the demand and supply within the MMG system. The net power requirement is positive (negative) if the sum of the MGO demand (supply) bids exceeds the sum of the MGO supply (demand) bids. Depending on the scenario, the PSO imports (exports) electricity from (to) the main grid.
\begin{subequations}
	\begin{align}
	&C^{m,t}_{{\rm PSO},h}\hspace{-1mm}\left(P^{m,t}_{{\rm PSO,b},h},P^{m,t}_{{\rm PSO,s},h}\right)\hspace{-1mm}=\beta^t_{{\rm b},h}P^{m,t}_{{\rm PSO,b},h}-\beta^t_{{\rm s},h}P^{m,t}_{{\rm PSO,s},h} \label{Cost_PSO}\\
	& 0 \leq P^{m,t}_{{\rm PSO,b},h},\,P^{m,t}_{{\rm PSO,s},h} \leq (P^{m}_{\rm PSO})^{\rm max};\ \forall h \in \mathcal{H} \label{PSO_ub1}
	\end{align}
\end{subequations}
where $ P^{m,t}_{{\rm PSO,b},h} $ is the power purchased by MG~$ m $ from the PSO at $ \beta^t_{{\rm b},h} $ and $ P^{m,t}_{{\rm PSO,s},h} $ is the power sold by MG~$ m $ to the PSO at $ \beta^t_{{\rm s},h} $. As explained later in Section~\ref{MarketClearing}, $ \beta^t_{{\rm b},h} > \beta^t_{{\rm s},h} $ 
which convexifies the power trading cost function in~\eqref{Cost_PSO} and avoids the simultaneous buying and selling of electricity. The constraints \eqref{PSO_ub1} bound the quantity of electricity imported from and exported to the main grid respectively.
\vspace{-3mm}%
\subsection{Energy Management Model for a Single Microgrid} \label{OpModel}
The total operating cost function under scenario~$ h $ during hour~$ t $ is formulated as follows.
\begin{align}\label{Cost}
&C_h^{m,t}\left(\Phi^{m,t}_h\right) =C^{m,t}_{{\rm G},g,h} +C^{m,t}_{{\rm E},e,h} +C^{m,t}_{{\rm FD},h} +C^{m,t}_{{\rm PSO},h}
\end{align}%
where $ \Phi^{m,t}_h $ is the decision variable vector pertaining to the local optimal dispatch problem (as shown below in~\eqref{MG_ED}) of MG~$ m $ under scenario~$ h $ during hour~$ t $.
To formulate the total expected cost function over the entire optimization horizon, the following assumptions are made on the basis of previous works in the literature~\cite{NianLiu17}.
\begin{enumerate}
	\item The probability of occurrence of the worst-case scenario is much lower than that of the base-case scenario. In this context, it is assumed that the MGOs know a potential value for the base-case probability~$ p_0 $ based on their prior experience.
	\item The \emph{sampling average approximation} method is applied in~\eqref{C_res}, wherein the probabilities of occurrences of scenarios $ h \in \mathcal{H}\backslash \{0\} $ during each hour are considered to be equal.
\end{enumerate}
Accordingly, the energy cost $ C^m_{\rm E} $ and the reserve cost $ C^m_{\rm R} $ for the robust operation of MG~$ m $ can be formulated as shown below in~\eqref{EnR_cost}.
\begin{subequations} \label{EnR_cost}
	\begin{align}
&C^m_{\rm E}=\sum_{t \in \mathcal{T}}C^{m,t}_{0}(\Phi^{m,t}_0)\\
&C^m_{\rm R}=\sum_{t \in \mathcal{T}}\left[\frac{1}{H}\sum_{h \in \mathcal{H}\backslash \{0\}}C^{m,t}_{h}(\Phi^{m,t}_h)-C^{m,t}_{0}(\Phi^{m,t}_0)\right] \label{C_res}
\end{align}
\end{subequations}
The value of $ p_0 $ can be selected on the basis of historical data and has no impact on the worst-case feasibility of the problem. In this context, the robust day-ahead optimal dispatch problem of each MG $ m \in \mathcal{M} $ can be formulated as shown below in~\eqref{MG_ED} to obtain the optimal \emph{expected} operation cost while respecting all the technical constraints under the worst-case scenario.
\begin{subequations}\label{MG_ED}
	\begin{align}
	\min_{\Phi^{m,t}_h} \ &C^m_{\rm exp}=C^m_{\rm E} + (1-p_0)C^m_{\rm R} \label{OBJ}\\
	{\rm s.t.}\ &\sum_{g \in \mathcal{G}^m}P^{m,t}_{{\rm G},g,h}+\sum_{e \in \mathcal{E}^m}\left[ P^{{\rm d},m,t}_{{\rm E},e,h}-P^{{\rm c},m,t}_{{\rm E},e,h}\right] +P^{m,t}_{{\rm PSO,b},h}\nonumber \\
	&\hspace{0mm}-P^{m,t}_{{\rm PSO,s},h}=P^{m,t}_{{\rm D},h}-\sum_{w \in \mathcal{W}^m}P^{m,t}_{{\rm W},w,h}-\sum_{s \in \mathcal{S}^m}P^{m,t}_{{\rm PV},s,h} ;\nonumber \\ 
	&\hspace{35mm}\forall h \in \mathcal{H},\forall t \in \mathcal{T} \label{MG_ED1}\\
	& \eqref{Cap_DG1},\, \eqref{ResG},\, \eqref{lin_DG},\, \eqref{Ech1}{\text -} \eqref{SOC},\, \eqref{Cap_E2}{\text -} \eqref{E01},\, \nonumber\\
	&\eqref{lin_ESS},\,\eqref{D}{\text -} \eqref{FD},\, \eqref{lin_D},\eqref{PSO_ub1};\,
	\forall t \in \mathcal{T} \label{MG_ED2}
	\end{align}
\end{subequations}

Here,~\eqref{OBJ} is the expected total cost function, \eqref{MG_ED1} is the power balance constraint, and \eqref{MG_ED2} consolidates the technical constraints explained earlier. 
The optimization problem in~\eqref{MG_ED} is a quadratic programming (QP) problem which can be solved efficiently using convex optimization solvers. Essentially, in \eqref{OBJ}, the behaviour of the base-case probability $ p_0 $ is similar to a weight factor which represents the trade-off between the base-case (higher $ p_0 $) and the worst-case (lesser $ p_0 $) cost functions. This improves the operational security and the economic efficiency of the MG.
\vspace{-2mm}%
\section{Optimal Energy and Reserve Sharing via Cooperative Bidding}

%
\subsection{Distributed Algorithm for Market Clearing}\label{MarketClearing}
The centralized optimization problem (COP) for the MMG system aggregates the sub-problems of the form in~\eqref{MG_ED} for all the MGs constituting the MMG system. The COP turns out to be a convex QP problem having a \emph{unique} optimum owing to the convex characteristics of~\eqref{MG_ED}. However, solving the COP requires the full system information of all the MGs to be known. In practice, individual MGs in the MMG system may be owned by separate entities. As such, these entities may be reluctant to share information such as generation costs and quantities owing to privacy requirements. Furthermore, simultaneously accessing large amounts of data may lead to data traffic issues. 
\begin{algorithm}[h!]%
	\caption{Cooperative Optimal Power Sharing Coordination}\label{Algo1}%
	\begin{algorithmic}[1]%
		\STATE \textbf{Initialize}  $ k=0 $, $\epsilon=0.005$, $ \lambda^t_{{\rm eq},h}=\lambda^t_{{\rm}h}(0) $ and $ \Delta P^t_{{\rm N},h}(0) $.
		\WHILE{$\underset{t \in \mathcal{T},h \in \mathcal{H}}{\max}\left| \Delta P^t_{{\rm PSO},h}(k)-\Delta P^t_{{\rm N},h}(k)\right| \geq \epsilon$}
		\STATE $ k=k+1 $ 
		\STATE Update $ \beta^t_{{\rm b},h} $ and $ \beta^t_{{\rm s},h} $ using~\eqref{MCP}. 
%
		\STATE \text{Extract power exchange preferences $ P^{m,t}_{{\rm PSO,b},h}(k)$}\\ \text{  $/P^{m,t}_{{\rm PSO,s},h}(k) $ from each MG; $ \forall t \in \mathcal{T} $, $ \forall h \in \mathcal{H} $.} \label{S5}
		\STATE \text{Compute $ \lambda^t_h(k+1) $ using~\eqref{lambda_eq}.} \label{S6}
%
		\IF{$  \lambda^t_h(k+1) \geq {\gamma}^{t}_{{\rm b},h}$} 
		\label{S7}
		\STATE $ \lambda^t_{{\rm eq},h} = {\gamma}^{t}_{{\rm b},h},\, P^t_{{\rm N, b},h} = \Delta P^t_{{\rm PSO},h}(k) $ \label{S9}
		\ELSIF{$  \lambda^t_h(k+1) \leq {\gamma}^{t}_{{\rm s},h} $} 
		\label{S10}
		\STATE $ \lambda^t_{{\rm eq},h} = {\gamma}^{t}_{{\rm s},h},\, P^t_{{\rm N, s},h} = -\Delta P^t_{{\rm PSO},h}(k) $ \label{S12}
		\ELSE \label{S13}
		\STATE $ \lambda^t_{{\rm eq},h} = \lambda^t_h(k+1),\, P^t_{{\rm N, b},h} = P^t_{{\rm N, s},h} =0 $ \label{S15}
		\ENDIF \label{S16}
		\STATE $\Delta P^t_{{\rm N},h}(k) =P^t_{{\rm N, b},h}-P^t_{{\rm N, s},h} $
		\ENDWHILE
	\end{algorithmic}%
\end{algorithm}
Algorithm~\ref{Algo1} summarizes the proposed cooperative bidding scheme proposed in this paper which mitigates the aforementioned issues. 
Therein, the COP is decomposed into $ M $ local sub-problems of the form in~\eqref{MG_ED}. All the constituent MGs make bids to the PSO on the basis of their power sharing preferences $ P^{m,t}_{{\rm PSO,b},h}/P^{m,t}_{{\rm PSO,s},h} $ under all the scenarios $ h \in \mathcal{H} $ over the optimization horizon $ t \in \mathcal{T}  $. The power sharing preferences are computed by executing \eqref{MG_ED} as described in Step~\ref{S5} of Algorithm~\ref{Algo1}. Subsequently, the PSO estimates the difference between the energy demand and supply bids to recalculate the market equilibrium price (MEP) $ \lambda^t_{{\rm eq},h} $ using \eqref{lambda_eq} as mentioned in Step~\ref{S6} of Algorithm~\ref{Algo1}. 
\begin{subequations}\label{lambda_eq}
	\begin{align}
	\Delta P^t_{{\rm PSO},h}(k) &= \sum_{m \in \mathcal{M}}\left[P^{m,t}_{{\rm PSO,b},h}(k) - P^{m,t}_{{\rm PSO,s},h}(k)\right]\\
	\lambda^t_h(k+1) &= \lambda^t_h(k)+\alpha\Delta P^t_{{\rm PSO},h}(k) \label{lambda_eq3}
	\end{align}
\end{subequations}
where $ k $ is the current iteration and $ \alpha \in \mathbb{R}_{>0}$ is the step-size. The power trading prices $ \beta^t_{{\rm b},h} $ and $ \beta^t_{{\rm s},h} $ for the MGOs are computed as in~\eqref{MCP} using the MEP $ \lambda^t_{{\rm eq},h} $ for scenario~$ h $ during hour~$ t $, wherein $ \tau $ is the per unit service fee imposed on both the buyer and seller MGOs.
	\begin{align}\label{MCP}
	\beta^t_{{\rm b},h}=\lambda^t_{{\rm eq},h}+\tau,\
	\beta^t_{{\rm s},h}=\lambda^t_{{\rm eq},h}-\tau
	\end{align}
The aggregated income of the PSO $ I^t_{{\rm PSO},h} $ under scenario~$ h $ at time~$ t $ in \eqref{PSO_Fee} should cover the power transfer losses and any possible expenses incurred for maintaining the system and facilitating the power sharing.

\begin{align}\label{PSO_Fee}
I^t_{{\rm PSO},h}=\tau \sum_{m \in \mathcal{M}}\left[ P^{m,t}_{{\rm PSO,b},h}+P^{m,t}_{{\rm PSO,s},h} \right] 
\end{align}

In this model, each MG~$ m $ solves \eqref{MG_ED} to minimize its generation cost for the given buying and selling trading prices represented by $ \beta^t_{{\rm b},h} $ and $ \beta^t_{{\rm s},h} $ respectively. In this respect, \eqref{lambda_eq} arrives at a higher MEP for the next iteration if the net demand is higher to encourage more suppliers or fewer buyers. Conversely, \eqref{lambda_eq} arrives at a lower MEP for the next iteration if the net supply is higher to encourage fewer suppliers or more buyers. Accordingly, \eqref{lambda_eq} describes a typical demand and supply market scenario. This can be interpreted as a \emph{market clearing} mechanism since the trading prices are modified till the net energy demand and supply are balanced~\cite{Gregoratti15}.

The MMG system in this paper is assumed to be operated in the grid-connected mode. The PSO manages the electric power transactions between the MMG system and main grid based on $ {\gamma}^{t}_{{\rm b},h} $ and $ {\gamma}^{t}_{{\rm s},h} $ which represent the per unit spot prices at which the electricity is purchased from and sold to the main grid respectively. In practice, $ {\gamma}^{t}_{{\rm b},h} > {\gamma}^{t}_{{\rm s},h} $~\cite{YunLiu18}. 
If the net supply bid is inadequate to meet the demand at $ \lambda^t_{{\rm eq},h}=\gamma^{t}_{{\rm b},h} $, the shortfall is met through imports from the main grid as described in Steps~\ref{S7}-\ref{S9} of Algorithm \ref{Algo1}, wherein $ P^t_{{\rm N, b},h} $ is the power purchased from the main grid. Similarly, if the net supply bid exceeds the demand at $ \lambda^t_{{\rm eq},h}=\gamma^{t}_{{\rm s},h} $, the surplus is exported to the main grid as described in Steps~\ref{S10}-\ref{S12} of Algorithm \ref{Algo1}, wherein $ P^t_{{\rm N, s},h} $ is the power sold to the main grid. 
Thus, $ \lambda^t_{{\rm eq},h} $ will lie in between $ \gamma^{t}_{{\rm b},h} $ and $ {\gamma}^{t}_{{\rm s},h} $ if the demand and supply bids can be internally matched as shown in Steps~\ref{S13}-\ref{S15} of Algorithm \ref{Algo1}.
%
\vspace{-3mm}%
\subsection{Convergence Characteristics}
The dual problem of the COP can be defined as follows.
\begin{subequations}\label{Dual}
\begin{align}
\max_{\lambda^t_{{\rm eq},h}} &\sum_{m \in \mathcal{M}} \min_{\Phi^{m,t}_h} C^m_{\rm exp}\left( \Phi^{m,t}_h,\lambda^t_{{\rm eq},h}\right) \\
{\rm s.t.} &\ \eqref{MG_ED1}, \eqref{MG_ED2};\ \forall m \in \mathcal{M}\\
&\ \underset{m \in \mathcal{M}}{\sum} \left[P^{m,t}_{{\rm PSO,b},h} - P^{m,t}_{{\rm PSO,s},h}\right]=P^t_{{\rm N, b},h}-P^t_{{\rm N, s},h}\label{CoupCon}
\end{align}
\end{subequations}

According to \eqref{Dual}, the local cost minimization subproblem \eqref{MG_ED} for each MG $ m $ is the contribution of the corresponding MG to the Lagrangian function at a given $ \lambda^t_{{\rm eq},h} $. Thus, $ {\boldsymbol \lambda_{\rm eq}}=\left[ \lambda^1_{{\rm eq},0},\ldots,\lambda^T_{{\rm eq},H}\right]^{\rm T}  $ comprises the Lagrangian multipliers $ \lambda^t_{{\rm eq},h} $ under scenario $ h $ during hour $ t $ relative to the coupling constraint~\eqref{CoupCon}. 
%
Algorithm~\ref{Algo1} is based on the \emph{iterative sub-gradient method} in~\cite[Ch. 8]{Bertsekas03} and derives the sequence $ \{{\boldsymbol \lambda_{\rm eq}}(k)\} $ which converges to the optimum of \eqref{Dual}. Each MG locally optimizes its contribution to the Lagrangian function at the given $ {\boldsymbol \lambda_{\rm eq}} $. Moreover, the aforementioned coupling constraints are interpreted in \eqref{lambda_eq} and in Steps~\ref{S6}-\ref{S16} of Algorithm~\ref{Algo1} as sub-gradients of \eqref{Dual} at $ \lambda^t_{{\rm eq},h} $. As such, $ {\boldsymbol \lambda_{\rm eq}} $ is sequentially updated in a direction which satisfies the coupling constraints. 
The \emph{zero duality gap} in convex optimization problems ensures the uniqueness and the global optimality of the solution obtained for \eqref{Dual}. Hence, the convergence of the sequence $ \{{\boldsymbol \lambda_{\rm eq}}(k)\} $ is guaranteed within a finite number of iterations for iterative sub-gradient methods depending on the selected step-size~$ \alpha $ in~\eqref{lambda_eq}~\cite{Gregoratti15}.

\vspace{-2mm}
\section{Case Studies}\label{Results}
\subsection{Simulation Settings}
Numerical simulations are performed on an MMG system comprising four MGs. Each MG contains one DG. The linear cost coefficients of the DGs in the MMG system are listed in Table~\ref{MG_Param}. Furthermore, for all $ m \in \mathcal{M}$ and for all $ g \in \mathcal{G}^{m} $: $ a^m_{{\rm G},g} =\$5{\rm /MW^2h} $, $ (P^{m}_{{\rm G},g})^{\rm max}  = 1\,{\rm MW} $, $ (P^{m}_{{\rm G},g})^{\rm min}  = 0\,{\rm MW} $, $ R^{{\rm up},m}_{{\rm G},g}=R^{{\rm dn},m}_{{\rm G},g}=0.5\,{\rm MW/h} $ and $ r^{{\rm up},m}_{{\rm G},g}=r^{{\rm dn},m}_{{\rm G},g}=0.3\,{\rm MW} $. ESSs are included in MG~$ 2 $ and $ 3 $ with the following parameters: For all $ m \in \{2,\,3\} $ and for all $ e \in \mathcal{E}^{m} $: $ a^m_{{\rm E},e} = {\rm\$ 5/MW^2h}$, $ B^m_e = {\rm 2\,MWh}$, $ \eta^{\rm c}_{e}= \eta^{\rm d}_{e}=0.97$, $ (P^{{\rm c},m}_{{\rm E},e})^{\rm max}=(P^{{\rm d},m}_{{\rm E},e})^{\rm max} ={\rm 0.8\,MW}$, $ (E^{m}_{e})^{\rm max}= {\rm 0.9\,}$, $ (E^{m}_{e})^{\rm min}= {\rm 0.1\,}$ and $ {E}^{m}_{{\rm R},e}= {\rm 0.4\,} $. Furthermore, MGs 1 and 4 contain dispatchable loads with the following parameters for all $ m \in \{1,\,4\}$: $ a^m_{\rm FD}= {\rm\$ 20/MW^2h}$, $ (P^{m,t}_{\rm RD})^{\rm max}=(P^{m,t}_{\rm CD})^{\rm max}= {\rm 0.4\,MW}$ and $ E^m_{\rm Shed}={\rm 1\,MWh} $.%
\renewcommand{\arraystretch}{1.2}%
\begin{table}[ht!]%
	\centering
	\vspace{-4mm}
	\caption{\footnotesize Linear Cost Parameters of DGs}
	\label{MG_Param}
	\vspace{-3mm}
	\begin{tabular}{c|c c c c}
		\hline
			Parameter	& MG $ 1 $ & MG $ 2 $ & MG $ 3 $ & MG $ 4 $\\
		\hline
		$ b^m_{{\rm G},g} \ {(\rm \$/MWh)}$ &  $ 90 $ & $ 70 $ & $ 80 $ & $ 100 $\\
		\hline
	\end{tabular}\vspace{-1.0em}
\end{table}%

\begin{figure}[!ht]%
	\begin{center}%
		\centering
		\vspace{-2mm}
		\includegraphics[width=3.2in]{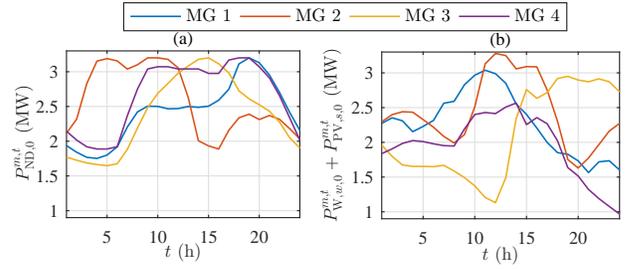} \vspace{-0.25cm}
		\caption{Forecasted (a) power demand and (b) aggregated RES generation.}\vspace{-0.7cm}
		\label{Load_RES}%
	\end{center}%
\end{figure}

The forecasts for the load demand and the RES generation are shown in Fig.~\ref{Load_RES}. Symmetrical forecast errors of $ [10\%,-10\%] $ and $ [5\%,-5\%] $ are considered for the nominal demand of each MG and the main grid spot prices respectively. The RES generation forecast errors are considered to be $ [10\%,-20\%] $. Moreover, $(P^{m}_{\rm PSO})^{\max} =1.5\,{\rm MW}$ for all MGs $ m \in \mathcal{M}$ and $ \tau=\$5/{\rm MWh} $.
\vspace{-3mm}
\subsection{Convergence Analysis}
\begin{figure}[!ht]
	\begin{center}
		\centering
		\vspace{-2mm}
		\includegraphics[width=3.2in]{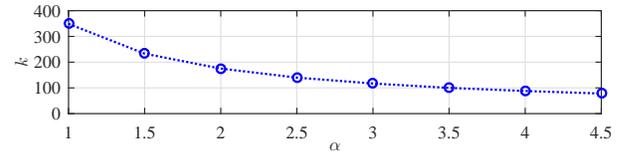} \vspace{-0.3cm}
		\caption{Number of iterations required for Algorithm \ref{Algo1} to converge for different values of~$ \alpha $.}\vspace{-0.6cm}
		\label{Alpha_fig}%
	\end{center}%
\end{figure}

\begin{figure}[!ht]%
	\begin{center}%
		\centering
		\vspace{-2mm}
		\includegraphics[width=3.2in]{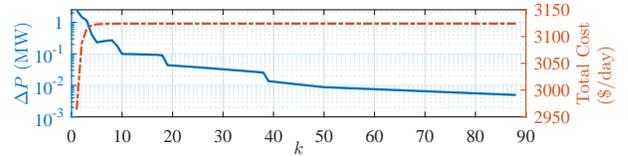} \vspace{-0.3cm}
		\caption{Convergence characteristics of Algorithm~\ref{Algo1}.}\vspace{-0.5cm}
		\label{Tol_fig}
	\end{center}%
\end{figure}

The convergence of Algorithm~\ref{Algo1} is influenced by the step-size~$ \alpha $ and the initial market equilibrium price~$ \lambda^t_h(0) $. Fig.~\ref{Alpha_fig} depicts the number of iterations required by Algorithm~\ref{Algo1} to converge to the optimal solution for different values of $ \alpha $. From Fig. \ref{Alpha_fig}, it is observed that Algorithm~\ref{Algo1} converges slowly when $ \alpha $ is smaller. However, the rate of convergence reduces exponentially when $ \alpha $ increases. The upper bound of $ \alpha $ was found to be approximately $ 4.5 $ in the simulation studies. After a few iterations, higher values of $ \alpha $ induced toggling in the power mismatch calculated as $ \Delta P=\underset{t \in \mathcal{T},h \in \mathcal{H}}{\max}\left| \Delta P^t_{{\rm PSO},h}(k)-\Delta P^t_{{\rm N},h}(k)\right| $.
Moreover, it was observed that $ \lambda^t_h(0) $ has minimal impact on the convergence of Algorithm \ref{Algo1}. Significantly, it was observed that the cooperative solution always converged to the centralized solution. Consequently, it can be concluded that the optimal solution attained using the cooperative trading scheme is independent of $ \alpha $ and $ \lambda^t_h(0) $. Fig.~\ref{Tol_fig} illustrates the evolution of the power mismatch and the objective function value with each iteration for $ \alpha=4 $, $ \lambda^t_h(0) = 0.5\big({\gamma}^{t}_{\rm b}+{\gamma}^{t}_{\rm s}\big)$ and $ p_0=0.5 $. From Fig. \ref{Tol_fig}, it is observed that the change in the objective function value is marginal after the first few iterations.%
\vspace{-4mm}%
\subsection{Robust and Deterministic Operations}%
\begin{figure}[!ht]%
	\begin{center}%
		\vspace{-0.4cm}
		\includegraphics[width=3.2in]{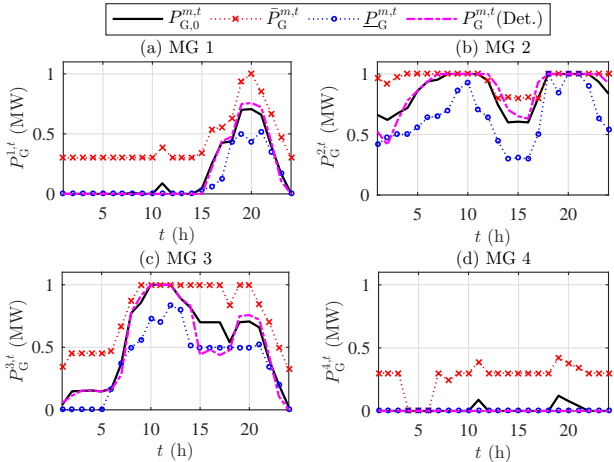}
		\vspace{-0.35cm}
		\caption{Power dispatch profiles for the deterministic and robust (base-case dispatch, and up and down reserve margins) operations of DGs in $ 4 $ MGs.}%
		\vspace{-0.5cm}%
		\label{Pg_fig}
	\end{center}%
\end{figure}

\begin{figure}[!ht]%
	\begin{center}%
		\vspace{-0.4cm}
		\includegraphics[width=3.2in]{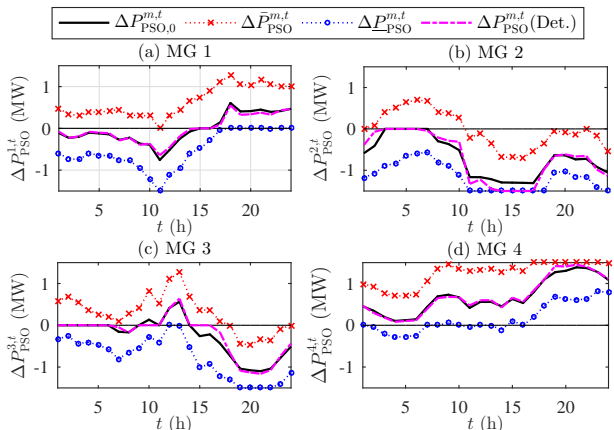}	\vspace{-0.35cm}
		\caption{Power and reserve exchanges between MGOs and the PSO.}\vspace{-0.6cm}%
		\label{MG_PSO_fig}%
	\end{center}%
\end{figure}

In this case study, the optimal dispatches of the MMG system using the deterministic (wherein the forecasts for the RES generation, demand and energy market prices are assumed to be $ 100\% $ accurate) and robust optimization approaches under Algorithm \ref{Algo1} are compared. 
Fig.~\ref{Pg_fig} shows the dispatches of the DGs in the MMG system at $ p_0=0.5 $ under the aforementioned simulation scenarios. In Fig. \ref{Pg_fig}, a higher utilization of the cheaper DGs is observed which signifies the effectiveness of the cooperative bidding scheme in optimally sharing cheaper resources across the individual MG boundaries. Furthermore, it is observed that the dispatches obtained using the RO and deterministic approaches differ. This is because the RO approach focuses on both optimal energy dispatch and reserve allocation while the deterministic approach focuses only on the optimal energy dispatch.
The solution obtained using the RO approach depends on $ p_0 $. 
Fig.~\ref{MG_PSO_fig} illustrates the trading of energy and reserves between the individual MGOs and the PSO. By and large, it is observed that MG~$ 2 $ behaves as a seller while MG~$ 4 $ behaves as a buyer. Conversely, MG~$ 1 $ and $ 3 $ alternate between the buyer and seller roles. Table~\ref{Cost_Rob_Det1} shows the energy and reserve costs of each MG and the PSO revenue for the RO and deterministic optimization approaches. 
The costs of the MGs using the RO approach exceed the costs using the deterministic approach. The cost of MG 4 is the highest since it houses the most expensive DG. Furthermore, a smaller $ p_0 $ prioritizes the cost of optimal reserve allocation in MGs $ 1 $-$ 3 $. The PSO income also follows a similar trend. This trend is not observed in MG $ 4 $ which is a buyer MG whose operational cost depends on the prices offered by MGs $ 1 $-$ 3 $ and the main grid.
\vspace{-3mm}%
\setlength{\tabcolsep}{5pt}
\renewcommand{\arraystretch}{1.2}%
\begin{table}[ht!]%
	\centering
	\caption{\footnotesize Cost Associated with Robust and Deterministic Operations}
	\label{Cost_Rob_Det1}
	\vspace{-2mm}
	\begin{tabular}{c|c|c|c|c|c}
		\hline
		Market & Deterministic & \multicolumn{4}{c}{Robust Operation $ (\$/{\rm day}) $} \\ \cline{3-6}
		Players & Operation & \multicolumn{2}{c|}{$ p_0=0.5 $} & \multicolumn{2}{c}{$ p_0=0.1 $}\\ \cline{3-6}
		& $ C^m_{\rm E}\,  (\$/{\rm day}) $ & $ C^m_{\rm E} $ & $ C^m_{\rm R} $ & $ C^m_{\rm E} $ & $ C^m_{\rm R} $\\
		\hline \hline
		MG $ 1 $ & $ 381.25 $ & $ 383.62 $ & $ 200.47 $ & $ 391.56 $ & $ 189.08 $\\
		MG $ 2 $ & $ 142.10 $ & $ 158.69 $ & $ 235.61 $ & $ 184.83 $ & $ 209.61 $\\
		MG $ 3 $ & $ 541.07 $ & $ 547.64 $ & $ 230.69 $ & $ 577.18 $ & $ 200.29 $\\
		MG $ 4 $ & $ 1598.28 $ & $ 1585.70 $ & $ 230.18 $ & $ 1566.52 $ & $ 240.22 $\\ 
		PSO & $ -239.71 $ & $ -244.35 $ & $ -56.83 $ & $ -249.10 $ & $ -49.28 $\\
		\hline
	\end{tabular}\vspace{-2.0em}
\end{table}%
%
%
\subsection{Comparison of Cooperative and Isolated Trading Schemes}
\begin{figure}[!ht]%
	\begin{center}%
		\vspace{-4mm}
		\includegraphics[width=3.2in]{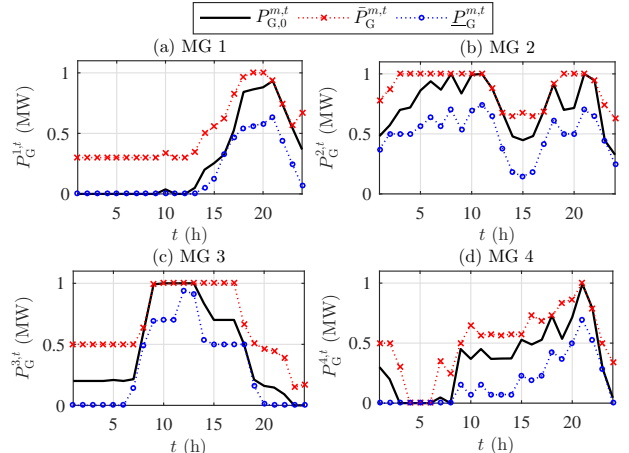} \vspace{-0.35cm}%
		\caption{Base-case power dispatch profile and up and down reserve margins of DGs under isolated trading in $ 4 $ MGs.}\vspace{-0.2cm}%
		\label{Pg_ISO_fig}%
	\end{center}%
\end{figure}

\begin{figure}[!ht]%
	\begin{center}%
		\vspace{-2mm}
		\includegraphics[width=3.2in]{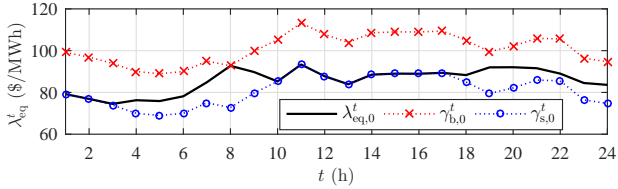} \vspace{-0.3cm}%
		\caption{Market equilibrium price for the deterministic operation and at the base-case of the robust operation.}\vspace{-0.5cm}%
		\label{lambda_fig}%
	\end{center}%
\end{figure}

\begin{figure}[!ht]%
	\begin{center}%
		\vspace{-1mm}
		\includegraphics[width=3.2in]{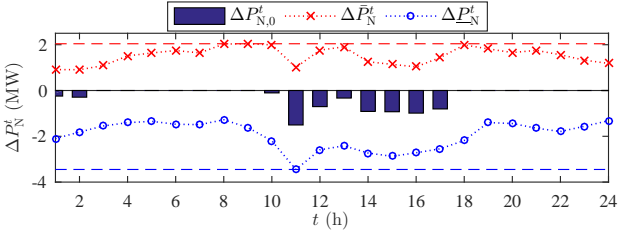} \vspace{-0.3cm}
		\caption{Power exchanges with the main grid and the reserve support expected. Contracted demand and supply are shown with red and blue dashed lines respectively.}\vspace{-0.7cm}%
		\label{Pgrid_fig}%
	\end{center}%
\end{figure}

In this case study, the performances of the cooperative and isolated trading schemes are compared for the robust operation of the MMG system. In the isolated trading scheme, it is assumed that the individual MGs trade with the main grid alone at the predetermined grid selling ($ \gamma^t_{\rm s} $) and buying ($ \gamma^t_{\rm b} $) prices. 
The dispatch profiles of all the DGs under the isolated trading scheme are shown in Fig.~\ref{Pg_ISO_fig} at $ p_0=0.5 $.
Compared with the cooperative trading scheme (dispatch profiles shown in Fig.~\ref{Pg_fig}), the utilization of the cheapest DGs in MGs~$ 2 $ and $ 3 $ under the isolated trading scheme is lower. From Fig.~\ref{Pg_ISO_fig}, it is seen that the most expensive DG in MG $ 4 $ is dispatched at full capacity during hour $ 21 $ while the relatively cheaper DG in MG $ 3 $ is partially dispatched.
In this scenario, the bidding framework embedded in the cooperative trading scheme allows MGs $ 2 $ and 3 to enhance their revenues while affording MG $ 4 $ an opportunity to reduce its expenditure.   
This can be observed in Figs.~\ref{lambda_fig} and \ref{Pgrid_fig} which illustrate the MEPs and the energy and reserve exchanges of the PSO with the main grid over the optimization horizon under the cooperative trading scheme respectively. Under the base-case scenario, the PSO only sells power to the main grid during the hours when the MEP settles at $ \gamma^t_{{\rm s},0} $. During the other hours, the MEP settles in between $ \gamma^t_{{\rm b},0} $ and $ \gamma^t_{{\rm s},0} $ owing to the cooperative trading scheme which accrues economic benefits to both the buyers and sellers when compared with the isolated trading scheme. Furthermore, the grid dependency of the MMG system is reduced, wherein the contracted demand (red dashed line in Fig~\ref{Pgrid_fig}) is $ 2.04\,{\rm MW} $ ($ 41.2\% $ reduction) and the contracted supply (blue dashed line in Fig~\ref{Pgrid_fig}) is $ 3.45\,{\rm MW} $ ($ 5.7\% $ reduction). 

\begin{figure}[!ht]%
	\begin{center}%
		\vspace{-2mm}
		\includegraphics[width=3.2in]{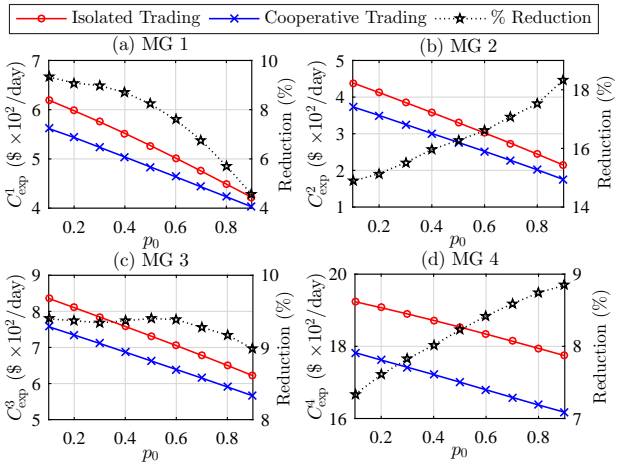} \vspace{-0.25cm}%
		\caption{Expected total operating cost in cooperative and isolated trading for $ 4 $ MGs along with the percentage cost reduction against~$ p_0 $.}%
		\vspace{-0.5cm}%
		\label{Cost_Opt_fig}%
	\end{center}%
\end{figure}

Finally, Fig.~\ref{Cost_Opt_fig} depicts the variations in the total operating cost incurred with~$ p_0 $ while using the RO approach under the cooperative and isolated trading schemes. A lower $ p_0 $ results in a higher reserve cost which leads to a higher total cost for each MG. 
Also, it is seen that the cooperative trading scheme enables the sharing of resources between the MGs thereby leading to a lower cost for all the MGs when compared with the isolated trading scheme. In this context, it is observed in Fig.~\ref{Cost_Opt_fig}(b) that the total cost of MG 2 reduces by 15-18\% under the cooperative trading scheme when the~$ p_0 $ varies.   
%
\vspace{-3.0mm}%
\section{Conclusions}
A cooperative trading scheme based on the iterative sub-gradient method was developed in this paper to enable the trading of energy and reserves between the four MGs constituting an exemplar MMG system. The MGs comprised DGs, ESSs, RESs, dispatchable loads and critical loads. The MG component models were formulated to comply with their respective intertemporal ramp constraints. These ramp constraints were taken into consideration while allocating the reserves. The TOAT method was used to identify a reduced set of vertices from the polyhedral uncertainty set which comprise the confidence bounds of the RES generation, the load demand and the energy market prices. Subsequently, a RO based approach was used to solve the optimal dispatch problem within each MG. The PSO was used to clear the MMG energy market based on the demand and supply bids submitted by individual MGOs. The results showcased the convergence characteristics of the cooperative trading scheme with respect to the step-size parameter of the iterative sub-gradient method. The efficacy of the cooperative trading scheme was demonstrated by comparing the total operating cost with that obtained by using the isolated trading scheme. The cooperative trading scheme resulted in economic benefits to the participating MGs owing to the shared utilization of cheaper resources.  
\vspace{-2mm}
\bibliographystyle{IEEEtr}
\bibliography{references_MMG}  
%

\end{document}